# Aligning tetracyanoplatinate thin films

Christian Rein[a, 1], Jens W. Andreasen[a], Martin M. Nielsen[b], Morten Christensen[b]

a. Department of Energy Conversion and Storage, Technical University of Denmark, 4000 Roskilde, Denmark.

b. Department of Physics, Technical University of Denmark, 2800, Kgs. Lyngby, Denmark.

CORRESPONDING AUTHOR

Christian Rein, Nanorein@gmail.com, DTU Energy, Technical University of Denmark, Frederiksborgvej 399, 4000 Roskilde, Denmark.

**Abbreviations:**

KCP, Potassium Tetracyanoplatinate(II); WAXS, Wide Angle X-ray Scattering; GIXD, Grazing Incidence X-ray Diffraction; AFM, Atomic Force Microscopy; FWHM, Full Width at Half Maximum.

# Abstract

By using a zone-casting derivative method it is possible to create well-aligned 200 nm thin films of tetracyano platinate crystal wires, which cover more than 50% of the surface of the substrate. The aligned crystal wires deviate only slightly from the casting direction and can each exceed 100 µm in length. Grazing incidence X-ray diffraction shows a 3.5 Å periodicity corresponding to the intrachain Pt-Pt distance found in single crystals and a 110 (crystal) orientation along the surface normal.

# I. INTRODUCTION:

Research on complexes containing potassium tetracyanoplatinate(II) (KCP) has been lying almost dormant for nearly 30 years [1,2]. However, as witnessed by recent reports on KCP-polymer composites and due to the increased performance of structural analytical tools, KCP research is now progressing from traditional single crystals and solutions towards thin films and composite materials [3,4,5]. When including crystal water into its structure, this compound crystallizes in the orthorhombic space group Pbcn, with unit cell axes (a,b,c) = 13.4 Å, 11.8 Å, 7.0 Å, and the complexes are lined up in chains along the c-axis [1]. Strong $d_z$-$d_z$ interactions between the Pt(II) centres of the KCP complexes leads to columnar stacking of the square planar units, ultimately resulting in needle shaped crystals containing parallel chains of platinum atoms. The possibility of doping the structure with e.g. bromine, makes KCP systems versatile and easy to manipulate through chemistry [6. Furthermore, doped KCP has been reported to display semi-metallic conductivity at room temperature [2]. The "free-electron"-like behaviour along the Pt-Pt chains would in theory lead to 1D quantization of the electron states – a property that opens up a new paradigm for electronics [7,8,9]. All these aspects of the KCP system, makes it an interesting candidate for future nanoelectronic and photonic applications, assuming that the system can be placed on a surface and controlled. Thus, in-depth structural and electronic information of KCP thin films will aid us in this endeavour, while simultaneously improving our understanding of the nanoscale surface phenomena of self-assembly.

Standard thin film fabrication methods like spin casting and drop casting of KCP solutions produce multi domain crystalline thin films with a large variation in topography [10]. In order to produce a thin film of aligned KCP-crystals, it is necessary to drive the crystallization along only one direction / crystal axis. This specific feature is found in zone casting and setups where sliding glass slides offer control of the thin film-forming solvent meniscus [11,12]. Previously, zone casting has been proven to enhance crystal alignment in thin films when compared to methods like spincasting and dropcasting [13]. We present here a derivative method of zone casting that allow easy alignment and fabrication of crystalline KCP thin films.

# II. EXPERIMENTAL DETAILS:

KCP stock solution of approx. 0.9 M was prepared by dissolving potassium tetracyanoplatinate(II) (KCP) (Sigma Aldrich) in Millipore grade water. Standard commercially available microscope glass slides (SuperFrost from Thermo Scientific, Green Color) were used as substrate and counter surface. Glass substrates for optical microscopy and AFM measurements were treated with oxygen plasma (1000 W, 10 min.), resulting in a super-hydrophilic substrate surface. The plasma surface treatment was found to significantly improve the alignment and coverage of the KCP thin film. It is likely that the improved wetting of the glass plates creates a more uniform shape of the solvent meniscus and thereby a more uniform concentration gradient and crystallization zone – as have previously been modeled for zone casting setups [11].

Samples were prepared using a brass stage with a brass lid as shown in figure 1. The counter surface slide was rigorously cleaned with ethanol and water before being placed on the stage with 5-8 μl KCP stock solution on top. The substrate slide was gently placed on top of the counter surface slide, such that the KCP solution was evenly distributed over the front half of the interface area. The stage lid was placed on top of the substrate slide and dry air was set to blow along the stage towards the interface filled with KCP solution. Stage lid and substrate were gently slid ¾ of the way along the stage with a speed of around 0.5-0.8 mm/s, before being lifted off for analysis.

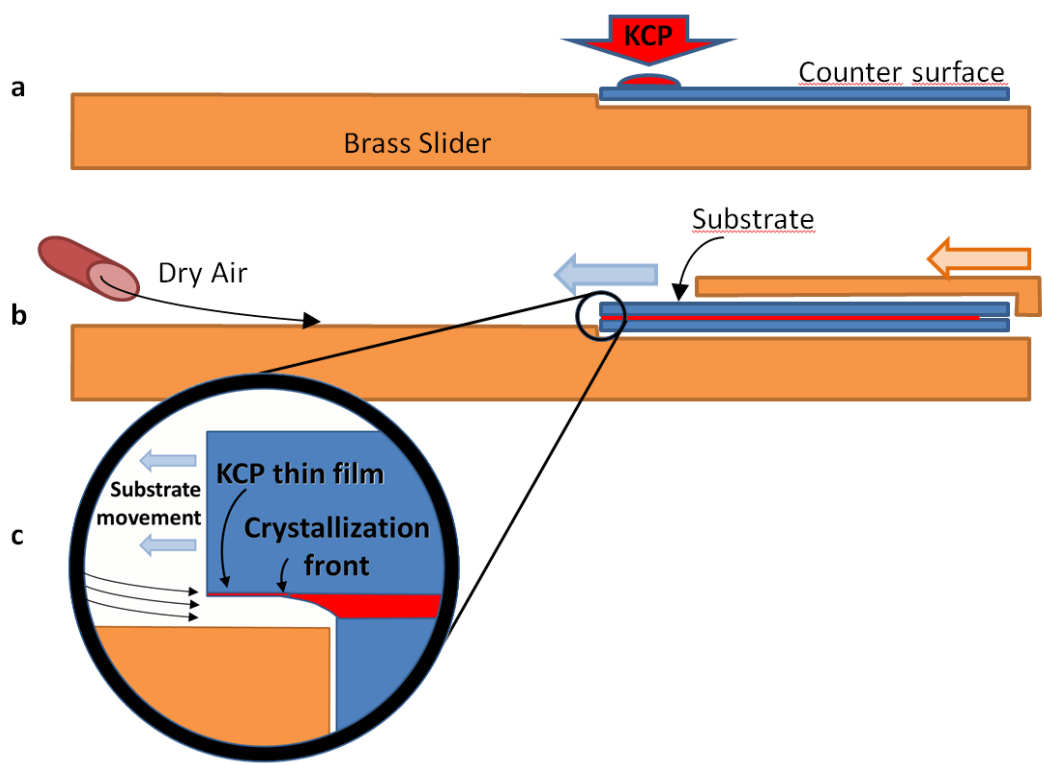

**Fig 1.** Illustration of the preparation of aligned crystalline thin films. a, counter surface on brass stage with KCP solution on top. b, super-hydrophilic substrate with stage lid is placed on top and dry air is set to flow along the stage. c, While moving the substrate the dry air concentrates the KCP solution and creates a crystalline thin film on the substrate.

Overall thin film quality was assessed with a Leica Optical Microscope (MZ16) together with a Leica CLS 150X light source. Optical imaging revealed an ultra-flat layer consisting of highly aligned crystal needles covering several square centimeters. The needles were parallel with the sliding direction.

A Nanosurf EasyScan atomic force microscope (contact mode at 20.47 nN force, commercially available probes from Olympus) was used to map the nanoscale topography at 5 different locations evenly distributed over the thin film. 70 μm x 70 μm images were recorded with the use of the Nanosurf EasyScan E-line program with a 0.6 s/line speed. Images were flattened and analyzed with Gwyddion [14]. Quantification of wire alignment in the AFM images was done by first masking the structures, based on height, and then correlating the maximum boundary size of each structure with the angle of the maximum boundary, relative to the sliding direction. Achieving a high degree of alignment would result in long structures with only small angular deviations. Application of a Gaussian fit to this correlation for each AFM image resulted in the factors a and b that describe both the size and alignment of the observed structures, respectively.

The surface-sensitive technique of grazing incidence X-ray diffraction (GIXD) was used for characterizing the KCP thin film in a wide angle X-ray scattering geometry (WAXS). The KCP thin film on glass was illuminated with the K-alpha X-radiation from a rotating copper anode, monochromatized with a bent, focusing multilayer optic and incident on the sample at a grazing angle of 0.2°. The scattering was collected on an X-ray sensitive imaging plate (Fuji BAS) for 45 min [15].

## III. RESULTS AND DISCUSSION:

Analysis of the AFM Images revealed a highly oriented system of crystal wires stretching longer than 70 µm (Width of AFM image) and a FWHM (Gaussian fit of 61-binned crystal length sum) of 5˚ (see figure 2). The preferred orientation of the structures observed in AFM corresponds to the optically observed orientation.

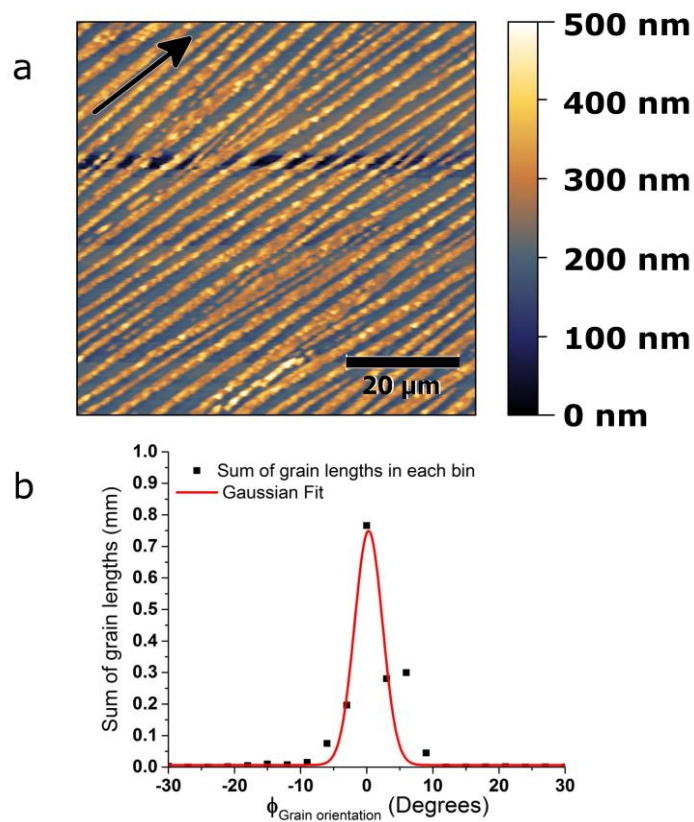

**Fig 2.** a, topographic image of a KCP thin film with an arrow to denote sliding direction. b, plotted correlation between 61-binned sum of grain lengths (crystal wires) and angle together with a Gaussian fit.

The resulting WAXS diffraction pattern of the KCP thin film (see figure 3a) shows an intense scattering in the plane of the substrate for the region around $Q_{xy} = \pm 1.8$ Å$^{-1}$ which corresponds to a periodicity along the surface of the substrate of approximately 3.5 Å. This distance compares well with the 3.47 Å Pt-Pt distance previously found in KCP single crystals [1] and gives us a strong indication of a preferred in-plane alignment of the platinum chains. Scattering features in the out-of-plane direction along $Q_z$ in the same region are groups of overlapping reflections, and signifies 3 dimensional ordering: intra- as well as interchain correlations resembling a microcrystalline packing [16].

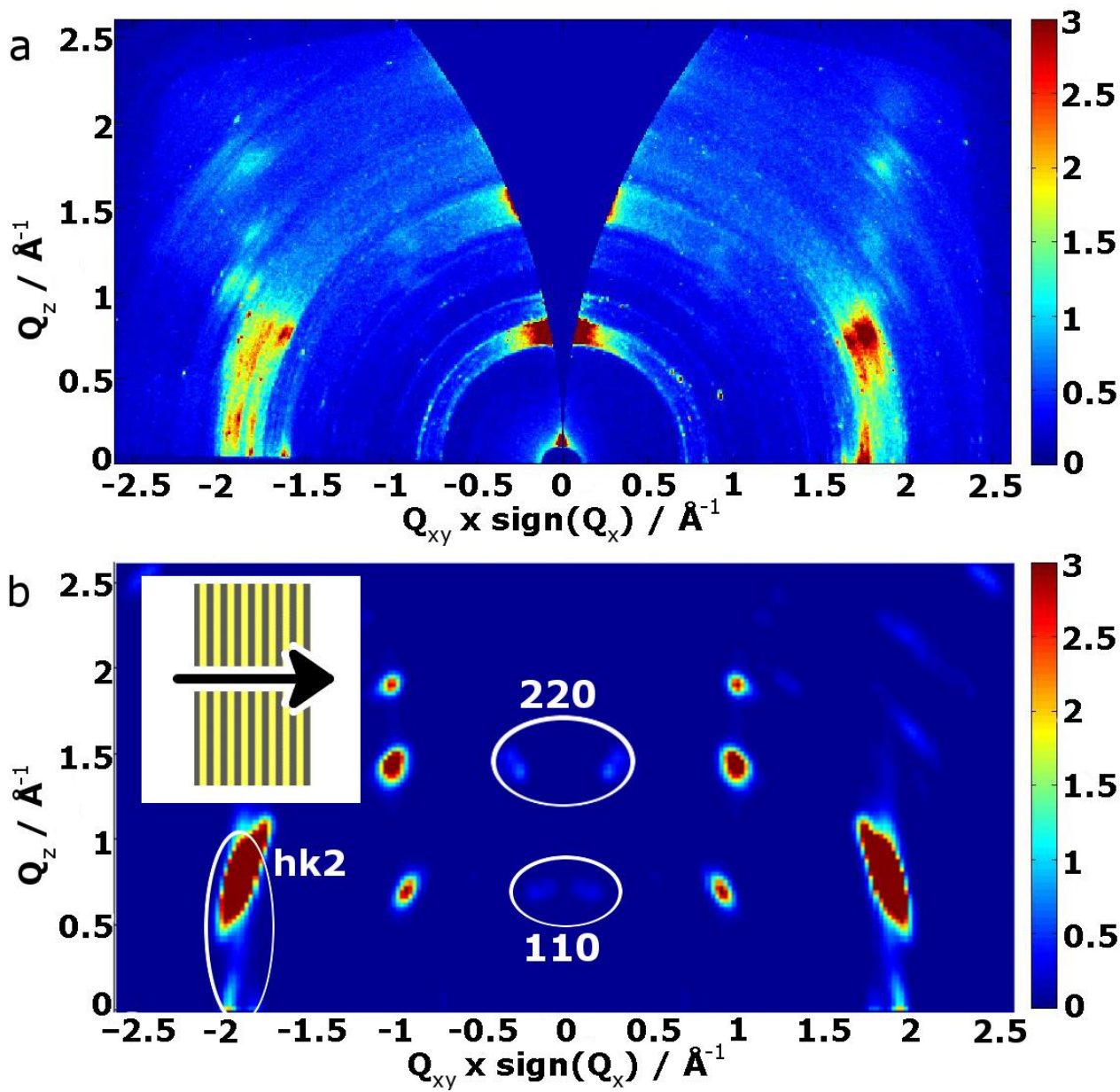

**Fig 3.** a, WAXS data for a KCP thin film on glass with crystal fibres oriented orthogonal to the incident X-ray beam. b, WAXS simulation of KCP primarily oriented with the 110-normal along the substrate normal (14° offset). Modelled scattering peaks assigned to hk2-planes (involving intrachain Pt-Pt distances) is observed as part of the group of overlapping reflections near $Q_{xy} = 2$ Å$^{-1}$. Insert, illustration of X-ray beam path relative to the orientation of the crystal fibres.

Comparing the first intense scattering on the $Q_z$-axis with that of the powder diffractogram of KCP shows good correlation with the 110 family of reflections and therefore indicates that the 110 plane of the crystal structure is predominantly oriented along the surface of the substrate. Taking this condition into account, a 2D WAXS diffraction pattern of KCP crystals on a surface was simulated (see figure 3b) [17, 18]. Both 110 and 220 scattering is observed in the model like in the data, but also the strong intensity of the hk2-peaks in the group of overlapping reflections match our model assuming a slight tilt of the unit cell corresponding to an angle between the 110 normal and the $Q_z$-axis of 14° . Whereas the simulation shows similar peak positions compared with our WAXS data, some peak intensities are different. Most noticeable is the very low intensity scattering from 02l, 20l and those higher orders peaks, which indicate a slight deviation of the surface-aligned crystals from the regular KCP bulk crystal structure.

The faint powder ring touching $Q_{xy} = 0.7$ Å$^{-1}$, $Q_z \sim 0$ Å$^{-1}$ shows that a small population of KCP crystals have a random orientation. The broad bands of scattering intensity in the radial direction are a result of the large X-ray beam footprint being imaged onto the detector, as scattering from different parts along the footprint results in a spread of effective sample-detector distances [17]. The extent of the substrate adds an uncertainty to the sample to detector distance, such that domains at one end of the substrate may occur at a slightly displaced scattering vector

The alignment of needle structures observed by optical microscopy was perpendicular to the incoming X-ray beam and the WAXS-data show that the Pt-Pt chains are oriented in the same direction as the needles. Previous measurements of emission polarization from oriented crystals of the Krogmann salt type in a polymer matrix have shown that the orientation of crystals and platinum chains coincide [3]. Our absorbance measurements using polarized light supports this model for the thin film systems (see Supporting Information).

## IV. CONCLUSIONS:

With the use of AFM it has been shown that application of a directional evaporation technique allows production of large scale thin films of aligned KCP crystals. Additionally, it was found that oxygen plasma pretreatment of the glass substrate improves distribution of the KCP solution prior to crystallization, which in turn improved the uniformity and alignment of the resulting KCP thin film crystals on the substrate. WAXS confirmed the in-plane alignment of platinum chains in the thin film and more work will be dedicated to improve our understanding of how self-assembled crystals adapt to surfaces.

## Acknowledgements:

Financial support from the Danish National Research Foundations Center for Molecular Movies is acknowledged. We would also like to acknowledge the great help and support from DTU

DANCHIP, Ole Trinhammer (Department of Physics, Technical University of Denmark.) and Erling Thyrhaug (Department of Chemistry, Copenhagen University).

**Supplementary Material for "Aligning tetracyanoplatinate thin films"**

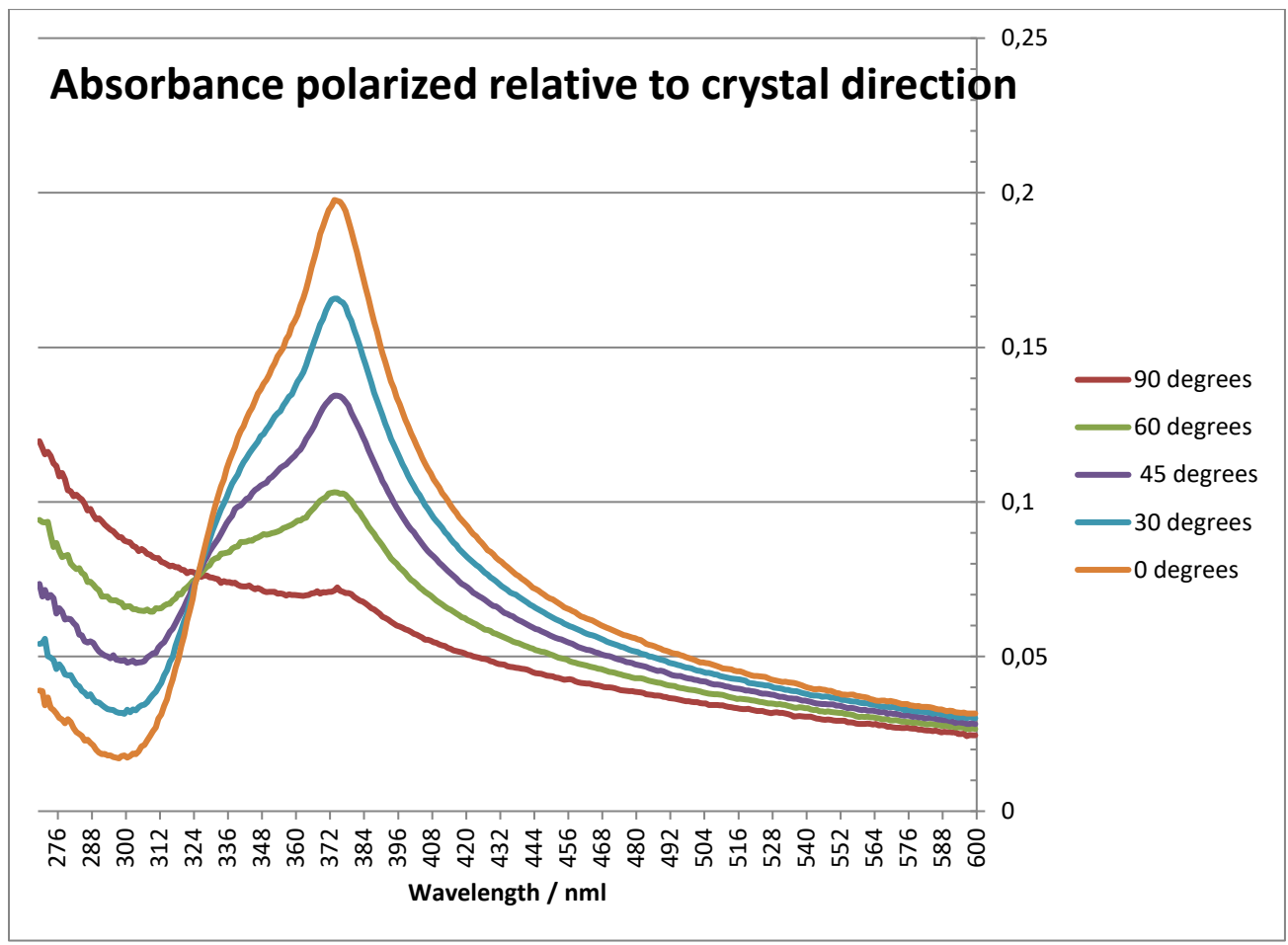